\documentclass[preprint,12pt]{elsarticle}

\usepackage{graphicx}
\usepackage{amssymb}
\usepackage{lineno, hyperref}
\usepackage{subcaption}
\journal{Clinical Oncology}

\begin{document}

\begin{frontmatter}

%% Title, authors and addresses

\title{Comparative analysis of radiotherapy LINAC downtime and failure modes in the UK, Nigeria and Botswana}

%% use the tnoteref command within \title for footnotes;
%% use the tnotetext command for the associated footnote;
%% use the fnref command within \author or \address for footnotes;
%% use the fntext command for the associated footnote;
%% use the corref command within \author for corresponding author footnotes;
%% use the cortext command for the associated footnote;
%% use the ead command for the email address,
%% and the form \ead[url] for the home page:
%%
%% \title{Title\tnoteref{label1}}
%% \tnotetext[label1]{}
%% \author{Name\corref{cor1}\fnref{label2}}
%% \ead{email address}
%% \ead[url]{home page}
%% \fntext[label2]{}
%% \cortext[cor1]{}
%% \address{Address\fnref{label3}}
%% \fntext[label3]{}

%% use optional labels to link authors explicitly to addresses:
%% \author[label1,label2]{<author name>}
%% \address[label1]{<address>}
%% \address[label2]{<address>}

\author[1]{Laurence M. Wroe}
\author[2]{Taofeeq A. Ige, Obinna C. Asogwa, Simeon C. Aruah}
\author[3]{Surbhi Grover}
\author[4]{Remigio Makufa}
\author[5]{Matthew Fitz-Gibbon}
\author[1]{Suzanne L. Sheehy\corref{cor1}}
\cortext[cor1]{Corresponding author} \ead{suzie.sheehy@physics.ox.ac.uk}
\address[1]{Department of Physics, University of Oxford}
\address[2]{National Hospital Abuja (NHA), Nigeria}
\address[3]{Department of Radiation Oncology, University of Pennsylvania. Botswana-UPENN Partnership}
\address[4]{Life Gaborone Private Hospital (GPH), Botswana}
\address[5]{Oxford University Hospitals NHS Foundation Trust}

\begin{keyword}
Radiotherapy \sep Accelerator \sep Failure Modes
%% keywords here, in the form: keyword \sep keyword

%% MSC codes here, in the form: \MSC code \sep code
%% or \MSC[2008] code \sep code (2000 is the default)

\end{keyword}

\clearpage
\begin{abstract}
%% Text of abstract
The lack of radiotherapy linear accelerators (LINACs) in Low- and Middle-Income Countries (LMICs) has been recognised as a major barrier to providing quality cancer care in these regions, along with a shortfall in the number of highly qualified personnel. It is expected that additional challenges will be faced in operating precise, high tech radiotherapy equipment in these environments, and anecdotal evidence suggests that LINACs have greater downtime and higher failure rates of components than their counterparts in High-Income Countries. To guide future developments such as the design of a LINAC tailored for use in LMIC environments, it is important to take a data-driven approach to any re-engineering of the technology. However, no detailed statistical data on LINAC downtime and failure modes has been previously collected or presented in the literature.

This work presents the first known comparative analysis of failure modes and downtime of current generation LINACs in radiotherapy centres, with the aim of determining any correlations between LINAC environment and performance. Logbooks kept by radiotherapy personnel on the operation of their LINAC were obtained and analysed from centres in Oxford (UK), Abuja, Benin, Enugu, Lagos, Sokoto (Nigeria) and Gaborone (Botswana). 

By deconstructing the LINAC into 12 different subsystems, it is found that the vacuum subsystem only fails in the LMIC centres and the failure rate in an LMIC environment is more than twice as large in 6 of the 12 subsystems compared to the High Income Country (HIC). Additionally, it is shown that despite accounting for only 3.4\% of total number of faults, the LINAC faults which take more than an hour to repair account for 74.6\% of the total downtime. The results of this study inform future attempts to mitigate the problems affecting LINACs in LMIC environments.

\end{abstract}

\end{frontmatter}
\clearpage 
%%
%% Start line numbering here if you want
%%
%\linenumbers

%% main text

\section{Introduction}
\label{S:1}

Radiation therapy is a critical component for treating and relieving the symptoms of cancer and is useful in half of all cancer cases \cite{HalfAll}. There is, however, a global disparity in the access to radiotherapy; in 2012, over 50\% of the approximately 4.0 million cancer patients in Low- and Middle-Income Countries (LMICs) who required radiotherapy were unable to access such treatment \cite{BenefitsOfExternalBeam}, \cite{SilentCrisis}. With many LMICs having inadequate or, in many cases, no radiation therapy centres, it is projected that to meet the LMIC radiotherapy demand over the next two to three decades, there is a need for around 12,600 radiation therapy machines \cite{ChangingRadTherParadigm}.

Radiotherapy can be delivered via a radioactive source, typically Cobalt-60, or by accelerating electrons in a linear accelerator (LINAC), producing X-rays by colliding the electron beam with a tungsten target. Although both technologies are mature and offer a range of benefits and drawbacks as a solution for providing external beam radiotherapy \cite{CompetingTechnologies}, it is argued by Coleman \textit{et al.} \cite{TreatmentNotTerror} that for reasons of security and safety, radiation delivered using a LINAC is the most effective solution to the radiotherapy burden in LMICs. Current generation LINACs, however, experience significant downtime in LMICs as they face challenges in these environments that they are not designed to manage. Their performance is adversely affected by regular interruptions to the energy supply, a lack of air temperature control in buildings and weak health care systems \cite{ExpandingGlobalAccess}.

Tackling the radiotherapy burden in LMICs is a complex task that requires multidisciplinary collaboration \cite{RadiotherapyRegionIncome}, \cite{Mobilising},  \cite{GlobalAccessWahab}. An International Cancer Expert Corps sponsored workshop held on the CERN campus in 2016 invited experts from fields including oncology and accelerator physics to consider future options, such as innovative technology, for tackling this global problem \cite{CERN2016Workshop}. The absence of detailed statistical data on LINAC downtime and failure modes, however, prevents the determination of the exact effect of the LMIC environment and its challenges on the performance of current LINAC technology. 

This work aims to quantitatively determine the effect of environment on LINAC performance. Failure mode data is obtained  and analysed from 14 current generation LINACs in the UK, Nigeria and Botswana; this sample offers variations in both socio-economic and physical environments and provides a dependent variable with which LINAC performance can be compared. The conclusions from this analysis allow for recommendations towards LINAC designs that are optimised for performance in challenging environments.

\section{Material and Methods}

\subsection{Collection and Sampling of LINAC Performance Data}

This study uses data obtained from 14 current generation LINACs: 6 from Oxford, UK, 6 from across Nigeria and 2 from Gaborone, Botswana, as detailed in Table \ref{T:1}. As the LINACs studied do not record or log their own performance data for local analysis, in order to analyse the LINAC failure modes and downtimes, data on machine performance was obtained from notes recorded by radiotherapists, medical physicists and engineers in logbooks at each institution. A typical entry would include the date and time of the fault, details on any interlocks and inhibits observed, how the LINAC was repaired and the amount of downtime the fault caused. There were, however, variations between the centres in both the level of detail in the description and whether  all faults were recorded or only the most severe; for instance, an average of 250 faults were recorded per LINAC per year in Oxford compared to just over 4 in Sokoto. In the case of incomplete information in the logbook, the most likely scenario was estimated based on other logbook entries and, where possible, this was made more accurate by liaising with the authors of the logbooks. It was assumed that all centres recorded all faults that caused more than an hour of downtime.

\begin{table*}[ht]
\centering
\begin{tabular}{p{1.7cm} p{1.4cm} p{5.95cm} p{3.2cm}}
\hline
& & & \textbf{Maximum}\\
& & & \textbf{Photon [MV] \&}\\
& & & \textbf{Electron [MeV]}\\
\textbf{Location} & \textbf{Comm.} & \textbf{Accessories} & \textbf{Energy}\\
%\textbf{Location} & \textbf{Comm*} & \textbf{Accessories} & \textbf{Maximum Photon [MV] \& Electron [MeV] Energy} \\
\hline
Oxford & 2007   &  MLC, kV, MV, PM & 15 \& 16 \\
Oxford & 2007   &  MLC, kV, MV, PM, VMAT & 15 \& 16\\
Oxford & 2007   &  MLC, MV, PM  & 15 \& 16 \\
Oxford & 2007   &  MLC, kV, MV, PM, VMAT  & 15 \& 16 \\
Oxford & 2007   &  MLC, kV, MV, VMAT, ExacTrac & 15 \& 16 \\
Oxford & 2007   &  MLC, MV, PM  & 15 \& 16 \\
\hline
Abuja & 1999  &   & 15 \& 18 \\
Abuja & 2017  & MLC & 15 \& 15 \\
Benin & 2013  &    & 15 \& 15 \\
Enugu & 2011  &    & 15 \& 15 \\
Lagos & 2009  &    & 15 \& 15 \\
Sokoto & 2009 &    & 15 \& 15 \\
\hline
Gaborone & 2001   &   & 10 \& 12 \\
Gaborone & 2015   & MLC, kV, MV, PM, VMAT  & 10 \& 12 \\
\hline
\end{tabular}
\caption{The sample of LINACs for which data was obtained for this study. Comm. column is the commissioning year, MLC (multileaf collimator), kV (kV imaging), MV (MV imaging), PM (patient position management), VMAT (volumetric modulated arc therapy). }
\label{T:1}
\end{table*}

This sample was chosen as there is a variation in environment with which LINAC downtime and failure modes can be compared, particularly in determining the effect of variation in socio-economic status (as of 2019, the World Bank classes the UK as an HIC, Botswana as an Upper-Middle Income Country and Nigeria as a Lower-Middle Income Country) and physical factors (for example the variation in climate and stability of power supply between the countries) on LINAC performance. There were also strong collaborative links to each of the centres. It should be noted that in this sample all HIC LINACs were from one vendor whereas all LMIC LINACs are from another. Additionally, the data available also covered different periods of the lifetime of the LINACs and so LINACs are not compared throughout the same stage of their life.

\subsection{LINAC Subsystems}

\begin{table}[h!]
\begin{subtable}{.5\textwidth}

\begin{tabular}{p{4.0cm} p{8.5cm}}
\hline
\textbf{Subsystem} & \textbf{Examples}\\
\hline
Air, Cooling$/$Gen & Generators, compressors, internal pipes, external chillers\\
Beam & Beam energy and symmetry\\
Computing & Monitors, keyboards, mice. (Does not include DICOM issues)\\
Couch and Door & Couch, tabletop, hand-pendant, external door for shielding\\
Diagnostics & Ionisation chamber\\
Gantry & Gantry timing belt, gantry bearings\\
Gun & Gun death, current issues, power supply issues\\
Positioning & Lasers, field lamps, position read outs (PROs), encoders\\
RF Power & Thyratron, klystron/magnetron, power cables\\
Shaping & Collimators, touch guard, carousel\\
Vacuum & Vacuum pumps, flanges, windows\\
MLC & MLC motors, MLC reflectors\\
& \\
\end{tabular}
%\caption{LINAC subsystems and examples.}
\end{subtable}

\begin{subtable}{.5\textwidth}
\begin{tabular}{p{3.5cm} p{9cm}}
\hline
\textbf{Fault Cause} & \textbf{Examples}\\
\hline
Mechanical & Switches, gearboxes, bearings, PROs, pipes \\
Electrical & Thyratron, fuses, poor electrical connections\\
Power & Power supply units, tripped circuit breakers, UPS\\
Board & PCBs, PSU boards, chips\\
Cabling & Power cables, signal cables\\
External & Generators, chillers, compressors, shielding door\\
Drift & Retuning of the beam\\
\hline
\end{tabular}
%\caption{LINAC fault causes and examples.}
\end{subtable}
\caption{Categories for a LINAC fault based on which subsystem failed and its cause.}
\label{T:2}
\end{table}

In order to compare the failure modes, rates and downtime of the radiotherapy machines between the different centres, the LINAC is deconstructed into 12 different subsystems and each fault is assigned 1 of 7 causes, as detailed in Table \ref{T:2}. Every entry recorded by the radiotherapy centre can be assigned to a subsystem and given an overall fault cause. By estimating the downtime for each centre, a failure rate per 1000 hours of uptime is calculated for each subsystem. The downtimes and failure rates of each subsystem are analysed in Section \ref{SS:1} to determine the effect of LINAC environment on performance.

%all subsystems detailed in~\cite{Karzmark2}

\section{Results}
\subsection{Analysis of Downtime in Oxford}
\label{SS:0}

 LINAC fault logs are typically concise notes recorded by relevant members of staff. Data is not currently recorded systematically enough to allow for automated analysis. To analyse the large amount of data available (i.e. 11875 faults recorded across 6 Oxford LINACs over a 7.5 year period), the dataset is sampled in two ways. 
 
 Firstly only faults affecting the LINAC and MLC, as detailed in Table \ref{T:2}, were analysed. The omitted systems included additional imaging systems (kV and MV), additional positioning and targeting (patient position management, respiratory gating), other systems (CT scanners) and communication and computing issues beyond those detailed in Table \ref{T:2} (DICOM). As the provision of these systems differs between environments, their omission from the analysis gives a more direct comparison of LINAC performance between centres. 

The second sampling technique was to only analyse the most severe faults that caused more than an hour of downtime. The justification is clear when binning the data according to its impact on downtime~\cite{Korean}, where $A<5$ mins, $5$ mins $<$ B $< 60$ mins, C$>60$ mins.

% \begin{itemize}
%     \item \textit{A} = Minor fault: Requires a quick reset and no investigation ($\leq$ 5 mins),
%     \item \textit{B} = Minor investigative fault: Typically requires an engineer to investigate the fault but little action or a minor fix required ($>$ 5 mins and $<$ 60 mins),
%     \item \textit{C} = Major fault: Failure of a component/system that causes significant downtime, requiring an engineer to perform a detailed investigation and then repair or fix the fault ($\geq$ 60 mins).
% \end{itemize}

\begin{table}[ht]
\centering
\begin{tabular}{p{1.5cm} p{1.7cm} p{1.3cm}  p{2.5cm} p{2.0cm} p{2.0cm}}
\hline
& & & \textbf{Total} & \textbf{Mean} & \textbf{Median}\\
%& & & \textbf{Downtime [hrs]} & \multicolumn{2}{c}{\textbf{Downtime [mins]}} \\
%& \textbf{Category} & \textbf{Faults} & \textbf{Downtime [hrs]} & \multicolumn{2}{c}{\textbf{Downtime [mins]}}\\
&  &  & \textbf{Downtime} & \textbf{Downtime} & \textbf{Downtime}\\
& \textbf{Category} & \textbf{Faults} & \textbf{[hrs]} & \textbf{[mins]} & \textbf{[mins]}\\
\hline
\textit{LINAC} & \textit{A} & 4122 & 119.1 (9.2\%) & 1.7 & 2\\
& \textit{B} & 666 & 210.4 (16.3\%) & 19.0 & 15\\
& \textit{C} & 171 & 965.4 (74.6\%) & 338.8 & 127\\
\hline
\textit{MLC} & \textit{A} & 588 & 21.7 (9.8\%) & 2.2 & 2\\
& \textit{B} & 408 & 96.6 (43.4\%) & 14.2 & 9\\
& \textit{C} & 23 & 104.0 (46.8\%) & 271.3 & 90\\
\hline
\end{tabular}
\caption{Comparison of failures in the LINAC and MLC systems of the Oxford data.}
\label{T:3}
\end{table}

Table \ref{T:3} shows that category \textit{C} faults account for 74.6\% of all downtime in the case of LINAC faults and 46.8\% of all downtime in the case of MLC faults. As category \textit{C} faults are the biggest contributors to downtime and reduce the data set to a size that allows for manual analysis, in this study we focus solely on these faults. 

Trends may exist between the occurrence of a more minor category \textit{A} or \textit{B} fault and the probability of a more severe category \textit{C} fault occurring in the near future and this may be useful in a study on preventative maintenance. However, as category \textit{A} and \textit{B} faults were not always recorded in LMIC centres, we are only able to compare C faults at present.

\subsection{An Overview of Downtime and Failure Rate Differences}
\label{SS:1}

The LINACs performances are grouped by country, but we refer to the UK as a HIC and Nigeria and Botswana as LMICs. To determine how the environment a LINAC operates in affects its performance, we analyse two factors: the downtime and rate of failure for each subsystem. The downtime is defined as the median downtime of each subsystem in each LINAC. Where more than one LINAC exists in a centre, the mean of these values is given.

The failure rate must be analysed with respect to the expected uptime of the LINAC. The uptime assumes a typical number of hours a LINAC would treat per week: based on discussions with the radiotherapy personnel, this is taken to be 50 hours every week for Oxford and 40 hours every week for all other centres. A superior measurement would be the failure rate per patients treated but differing patient loads and ethical considerations of patient records make this impractical at present.

The failure rate per hours of uptime (inversely proportional to the mean time between failure) is calculated by dividing the total number of category \textit{C} faults by the total uptime of each centre, which is displayed in Table \ref{T:4}. By dividing by the uptime of each centre, the contextual issues that influence LINAC availability do not affect the failure rate results.

However, contextual issues will necessarily affect downtime results, as centres spend different amounts of time waiting for parts and waiting for engineers, and centres have engineers of varying experience. The ways in which this contextual information affects conclusions is discussed in Section \ref{S:4}. 

\begin{table*}[ht]
\centering
\begin{tabular}{p{3.2cm}| p{2cm} p{2cm} p{2.0cm}| p{2.2cm}}
\hline
\textbf{Location} & \textbf{Data} & \textbf{Data} & \textbf{Hours of} &  \textbf{Calculated}\\
\textbf{(Commissioned)} & \textbf{Format} & \textbf{Covering} & \textbf{Data} & \textbf{Downtime}\\
\hline
Oxford (2007)   & e-Database & 2011-2018 & 19536 & 1.3\% \\
Oxford (2007)   & e-Database & 2011-2018 & 19536 & 0.7\% \\
Oxford (2007)   & e-Database & 2011-2018 & 19536 & 1.7\% \\
Oxford (2007)   & e-Database & 2011-2018 & 19536 & 1.4\% \\
Oxford (2007)   & e-Database & 2011-2018 & 19536 & 1.0\% \\
Oxford (2007)   & e-Database & 2011-2018 & 19536 & 0.7\% \\
Abuja (1999)    & Logbook & 2008-2017$\dagger$ & 17377 & 22.7\% \\
Abuja (2017)    & Logbook & 2017-2018 & 1909 & 4.2\% \\
Benin (2013)    & Logbook & 2013-2018 & 5640 & 14.2\% \\
Enugu (2011)    & Logbook & 2011-2018 & 14080 & 54.7\% \\
Lagos (2009)    & Logbook & 2009-2018 & 16720 & 18.8\% \\
Sokoto (2009)   & e-Logbook & 2009-2018 & 17423 & 20.0\%* \\
Gaborone (2001) & e-Logbook & 2001-2014$\dagger$ & 28343 & 1.3\% \\
Gaborone (2015) & e-Logbook & 2015-2018 & 4583 & 1.3\% \\
%Display - location (commission yeadata forr), mat, hours of data available, recorded downtime, operating hours per week, years of data, downtime estimate
\hline
\end{tabular}
\caption{Details on the data available on each LINAC and the calculated downtime ($\dagger$ denotes LINAC decommissioned, * denotes estimate).}
\label{T:4}
\end{table*}

\begin{figure}[h!]
\includegraphics[scale=0.50]{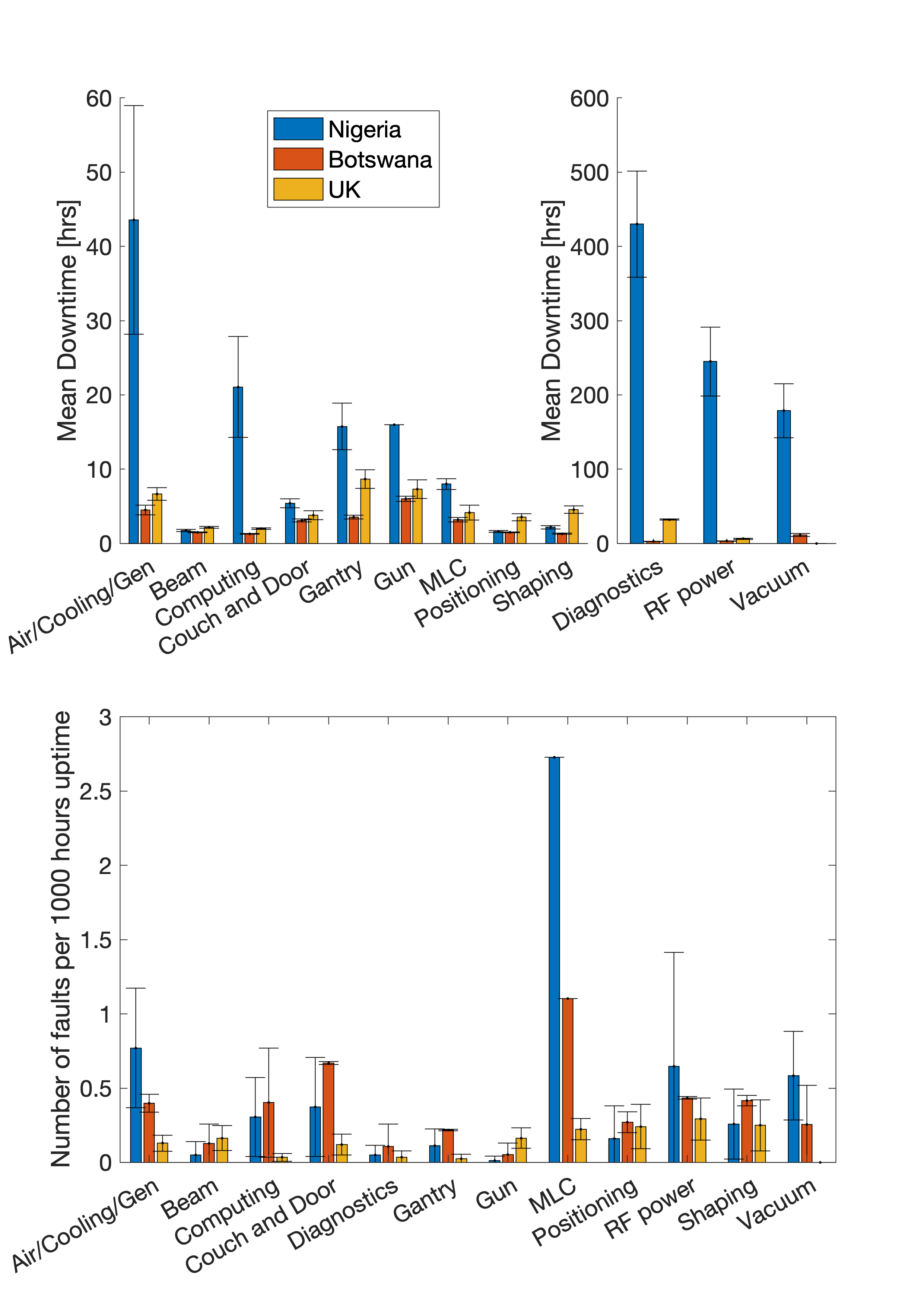}
\centering
\caption{(Upper) The mean downtime of faults caused by each subsystem. The error bars represent the standard deviation divided by 10 to give an indication of the spread. (Lower) The number of \textit{C} faults that occur in each country per 1000 hours of LINAC uptime. The error bars represent the standard deviation to give an indication of the spread. In both figures, the mean is calculated as the mean of the median downtime of each subcategory at each centre.}
\label{F:2}
\end{figure} 

Figure \ref{F:2} shows the mean downtime, as well as the failure rates of the LINAC subsystems per 1000 hours of uptime. The mean downtime in each subsystem is comparable between the Botswana and UK centres but significantly larger in Nigeria for the majority of subsystems. This appears to reflect the different service contracts the countries have in place: the Oxford centre has a full parts contract with the manufacturer, the Gaborone centre has a full parts and service contract with the manufacturer, whereas the Nigeria centres do not have either a parts or service contract. The mean downtime in Nigeria is significantly larger for the Diagnostics, RF Power and Vacuum subsystems. This is perhaps related to the significant cost to replace the ion chamber, thyratron/magnetron and ion pump. Such contextual issues are discussed further in Section \ref{S:4}. 

The failure rate is greater in LMIC environments for all subsystems except for the beam, positioning and gun. The air, cooling and generator and vacuum subsystems are discussed in detail in the following sections. Other results include:

\begin{itemize}
    \item{Computing: this subsystem fails more than 9 times as often in Nigeria and Botswana than Oxford. In the LMIC environments, computing equipment is not as readily available as in Oxford, so faults can be escalated to category \textit{C}, requiring complex repairs rather than a simpler, but more expensive, replacement.}
    
    \item Couch and external door: The failure rate in the LMIC environments is 3 times greater than in Oxford. This subsystem is affected by power cuts (and the subsequent surges when the power returns) causing fuses, that are not trivial to find or replace, to blow. There are more door issues in the LMIC due to their LINACs requiring a large mechanical door for shielding and safety purposes, whereas the infrared sensor systems used in Oxford appear to generate fewer issues. 
    
    \item RF power: this subsystem fails twice as often in Nigeria, however this result is skewed by an outlier from the Abuja (2017) LINAC (arising due to the thrice repeated failure of the 10A fuse in the thyratron pulse assembly in the short time data is available for) which significantly increases the mean downtime. LMIC centres have more RF power faults caused by power issues than HIC centres. 
    
    \item Gantry: there are 4 times more failures of this subsystem in LMICs. There are only 3 category \textit{C} gantry faults in the Oxford data and this small number may be due to more frequent planned maintenance of the gantry system in Oxford, quicker repairing of gantry faults (meaning a higher proportion of faults are category \textit{B} rather than \textit{C}) or this could be a vendor difference.
    
    \item MLC: Although the failure rate is at least 4 times greater in the LMIC environments, the fact that only the Oxford, Abuja (2017) and Gaborone (2015) LINACs have MLCs contributes to the large apparent disparity between the two environments. The data available for the two LMICs is from their date of installation, compared to 4 years after the installation of the LINACs in the HIC. As a result, the data in the LMICs may be skewed by `early failures'. Furthermore, the data available for the Abuja (2017) and Gaborone (2015) LINACs is small (1902 and 4583 hours respectively) and thus statistical fluctuations have a large impact on the calculated failure rate. To compare the performance of the MLC between environments, more data should be collected.
    
    \item Diagnostic: The rate of failure is comparable between the environments. The rate of failure of the ion chamber itself is very consistent between the environments, the slight difference is caused by more failures of the board equipment relating to the ion chamber in Gaborone.
\end{itemize}

A few subsystems appear to fail more frequently in the HIC than the LMICs; these are the beam, positioning and gun subsystems. 
\begin{itemize}
    \item Beam: The beam failure rate may be greater in the HIC because these issues are always recorded by Oxford, whereas they are not necessarily always recorded at other centres due to the nature of the issue.
    
    \item Positioning: The failure rate is slightly greater in the HIC data and this appears to be due to Oxford having a greater number of issues with their PROs (Position Read Out), SPROs (Secondary PRO) and encoders. This may result from vendor differences or tighter tolerances imposed in the Oxford centre.
    
    \item Gun: Higher HIC failure rate is most likely due to the difference in design between the vendors. In Oxford, the gun had 17 category \textit{C} faults across the 6 LINACs for issues requiring it to be re-potted or replaced. In contrast, the only comparable issue the gun had in the LMIC datasets was that it required replacing twice on the Gaborone (2001) LINAC. This highlights the importance in the design of the gun subsystem.
\end{itemize} 

\subsubsection{Air, Cooling and Generator Subsystem} \label{S:2}

\begin{figure}[h!]
\includegraphics[width=0.9\textwidth]{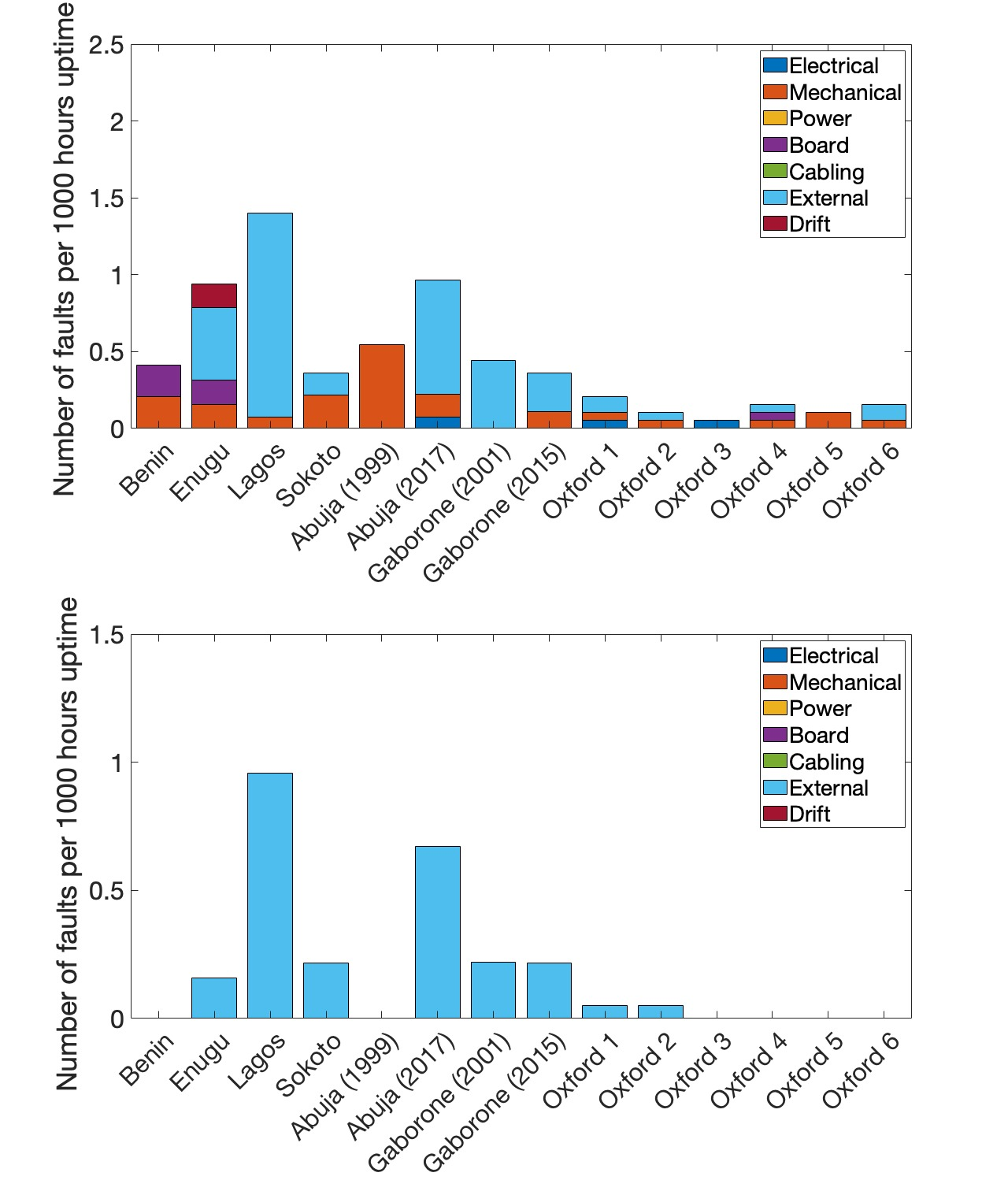}
\centering
\caption{(Upper) The number of air, cooling and generator subsystem category \textit{C} faults that occur in each centre per 1000 hours of uptime. (Lower) Comparison between category \textit{C} chiller failures at different centres.}
\label{F:3}
\end{figure}

As shown in Figure \ref{F:2}, air, cooling and generator faults are at least 3 times as frequent in the LMIC environments than in Oxford. Figure \ref{F:3} (upper) shows a breakdown of the failure rate of the different LINACs studied. The most prominent failures for this subsystem are mechanical and external failures. The mechanical failures are mostly leaking pipes and low water and gas pressures which cause significant downtime if the root cause of the issue cannot be determined. External failures result from issues with external chillers, generators and compressors (if present).

All centres have an external chiller, yet it is evident from Figure~\ref{F:3} (lower) that the chillers fail more often in LMIC environments, perhaps due to operating in a hotter, dustier environment. Active maintenance was observed at the Oxford centre with weekly checks and observations performed on the chiller by the local engineers; similar procedures at all centres could improve uptime. Chiller and generator maintenance is often subcontracted out in LMICs and thus the faults are not necessarily recorded in logbooks (which may explain the absence of reported chiller and generator faults in Benin).

The power supply differs between environments. The Benin, Enugu, Lagos and Sokoto centres are solely powered by generators to circumvent the frequent power cuts resulting from the instability of the grid power supply. In Abuja, the grid is used with a dedicated back up generator. The use of generators creates an additional single-point failure mode; if the generator is down (reported issues include running out of fuel and fires), so is the LINAC. 

Generators are likely to be necessary in LMIC environments in future, but their implementation needs careful planning to avoid extra downtime. This is evidenced in a study on radiotherapy in Botswana~\cite{StoryBotswana}, where there is a clear increase in unplanned downtime resulting from changing from the more stable power supply of South Africa to that of Botswana from 2012 onwards. The implications of power failure on the LINAC and recommendations for managing this is discussed further in Section \ref{S:3}.

\subsubsection{Vacuum Subsystem} \label{S:3}

\begin{figure}[ht]
\includegraphics[width=1.1\textwidth]{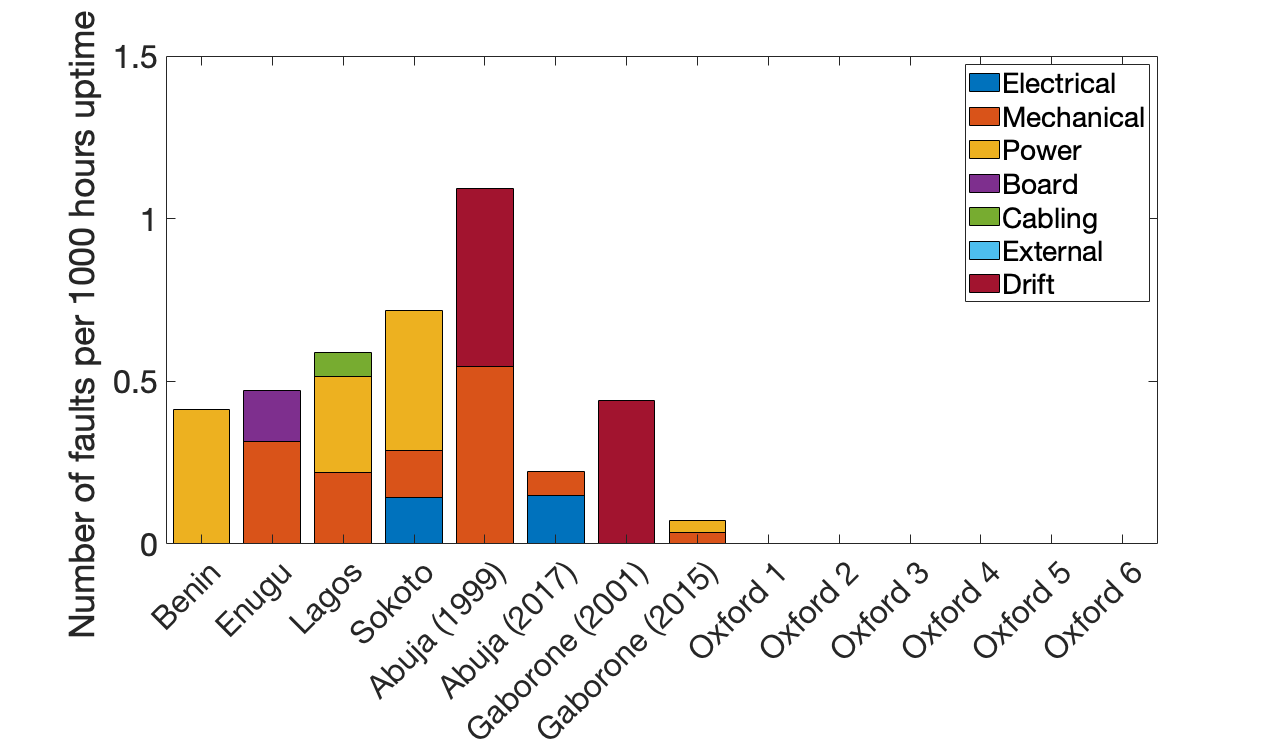}
\centering
\caption{The number of vacuum subsystem category \textit{C} faults that occur in each centre per 1000 hours of uptime.}
\label{F:5}
\end{figure} 

 Figure \ref{F:5} displays the failure rate of the vacuum subsystem and it is this subsystem that has the most striking difference between the HIC and LMIC environments. There are no recorded failures in any of the 6 HIC LINACs whereas there are recorded faults in all LINACs at the LMIC centres. The failure of the vacuum is not a trivial issue: depending on the amount of contamination, level of vacuum to recover and any damage to pumps, a failure can cause hours to weeks of downtime.

This is a clear environmental factor that is not experienced in HICs and affects the performance of the LINAC. The vacuum is susceptible to multiple routes of failure as a result of interruptions to the power supply. Irregular power supplies can affect the temperature regulation of the LINAC, leading to overheating and the vacuum pressure drifting. Power surges cause fuses to fail affecting many subsystems and components, including the ion pump. Finally, a common and dramatic failure mode is the loss of power to a backing pump, leaving a (poorly maintained) ion pump to support the vacuum. The ion pump fights a losing battle trying to keep the vacuum and eventually overheats and fails, causing a total loss of vacuum to atmosphere. The LINAC must then be brought back down to vacuum and the (expensive) ion pump must be replaced.

\subsection{Downtime in Enugu} \label{S:4}

\begin{figure}[ht]
\centering\includegraphics[width=1.0\linewidth]{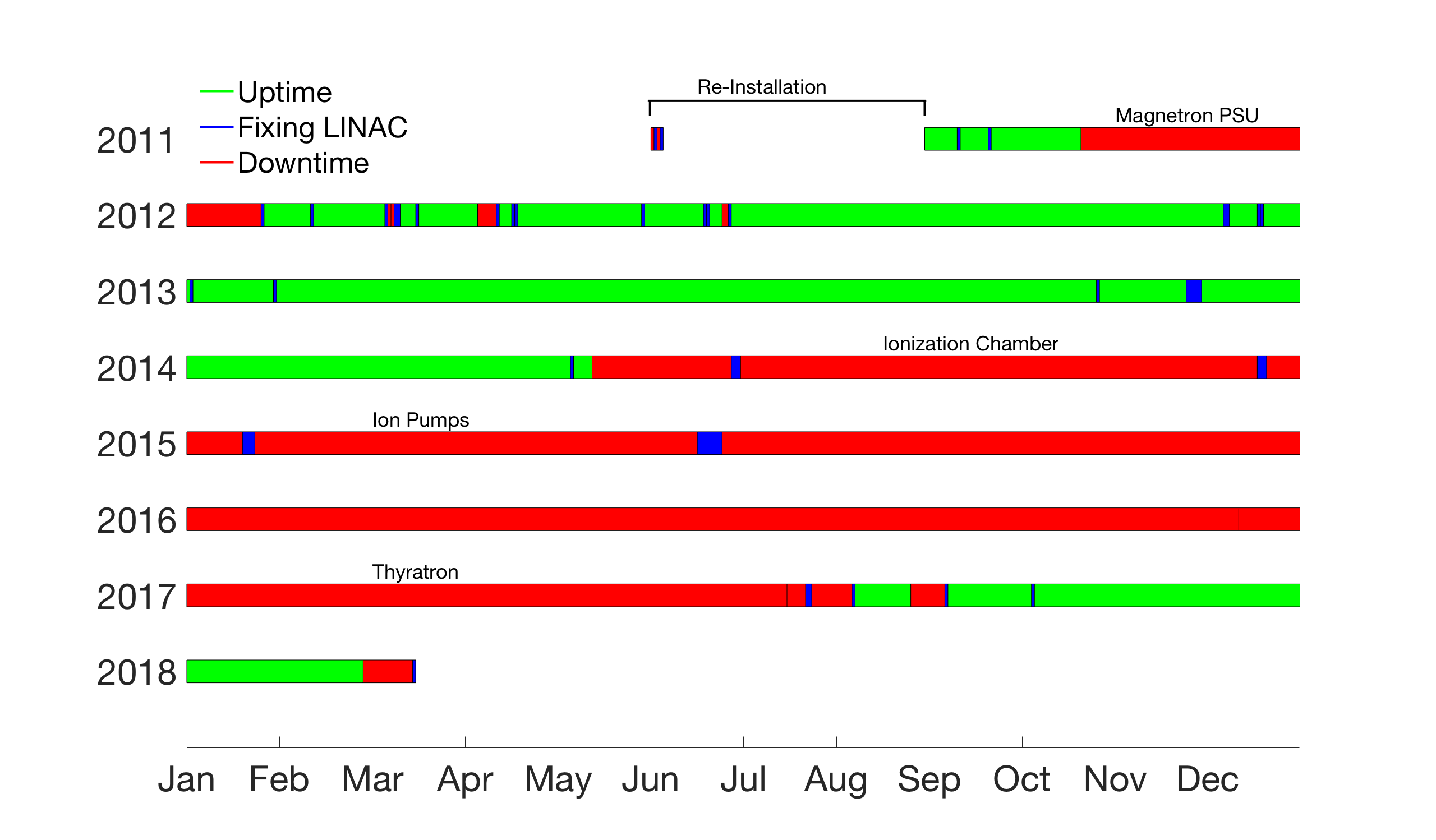}
\caption{Downtime for Enugu (2011) LINAC.}
\label{F:1}
\end{figure}

Contextual factors including the number and skill of local engineers, ease of access to spare parts and the level of contractual support available all affect the downtime of the LINAC. This is particularly evident in Figure \ref{F:1} which visually represents the 54.7\% downtime experienced by the Enugu (2011) LINAC. This figure agrees with the qualitative experiences of downtime discussed by Reichenvater \textit{et al.} \cite{Graveyard}.

The overall downtime is dominated by a few long periods rather than many frequent, small periods. The Enugu machine was initially installed in 2007 but a vandalisation (scavenging the system for valuable parts) and a fire delayed the machine treating patients to 2011. After the period of downtime from May 2014, the formation of a private-public partnership (PPP) in 2017 enabled the centre to start running again. The reasons for the long periods of downtime include:

\begin{itemize}
    \item Waiting for parts. It was reported by multiple LMIC centres without service contracts, including Enugu, that the administrative process of sourcing funds for spare parts can take so long that exchange rate fluctuations make quotes invalid. The whole, lengthy internal process must then repeat.
    \item Waiting for specialist engineers who can assist with troubleshooting and diagnosing a fault, or performing a complex repair. Local engineers may have difficulty troubleshooting LINAC failures because they have no experience in LINAC maintenance. Some centres cannot afford to send them on the vendor recommended training courses for LINAC engineers so they are trained `in-house' in the country. They also struggle to interpret the interlocks and inhibits reported by the machine when a fault occurs.
    \item When the machine has been down for a long time, patients are referred for treatment elsewhere. After repair, the centre must go through a lengthy administrative process for operating as a treatment centre again and this is not a trivial issue. 
\end{itemize}

\section{Discussion} \label{S:5}

The results of this paper are based on an analysis of logbooks and databases kept by radiotherapy personnel. The results obtained show a key failure rate difference between the vacuum and air, cooling and generator subsystems.  Maintaining vacuum during power shortages is critical. In Abuja, the local engineer has built a UPS that supports the vacuum during periods of power failure, and a similar system could be incorporated into the design of the machine. The LINAC should also be designed so that it shuts down safely when power is absent, with a passive valve to maintain vaccum and preserve the ion pump. A sealed vacuum unit that requires minimal to no pumping could be an excellent solution to solving the problem of maintaining the vacuum in periods of power outage. The results also show a reduced failure rate when the generators and chillers are regularly maintained and observed and this is recommended at all centres.

We note that differences in recording practises (severity of faults recorded, detail recorded and length and frequency of periods where no log is kept) mean that there is a systematic difference in the recording of faults. An improvement in logbook keeping practise would improve further studies.

\section{Conclusion}

This study presents an analysis and comparison of the performance of LINACs between different environments based on logbooks and databases. It is shown that failures of the air, cooling and generator, computing, couch and door, RF power and vacuum subsystems all appear to have significantly different rates of failure between HIC and LMIC environments and the underlying reasons for these different rates is discussed. Furthermore, it is shown that the reliance of LMICs on generators means that faults associated with the generators themselves make them a significant cause of failure. The unstable power supplies in LMICs can affect other subsystems, most notably the vacuum. Contextual issues are also discussed and how waiting for replacement parts, the skill and experience of local engineers and slow internal processes all have a very significant impact on LINAC downtime. Recommendations are made regarding design adjustments that could improve LINAC performance in LMICs.

\section*{Acknowledgements}
This work was supported by the Laidlaw Scholars Undergraduate Research \& Leadership Programme, Science and Technology Facilities Council [grant number ST/S000224/1] and The Royal Society [grant number UF160117].

%% References
%%
%% Following citation commands can be used in the body text:
%% Usage of \cite is as follows:
%%   \cite{key}          ==>>  [#]
%%   \cite[chap. 2]{key} ==>>  [#, chap. 2]
%%   \citet{key}         ==>>  Author [#]

%% References with bibTeX database:
\bibliographystyle{model3-num-names}
\bibliography{references}

%% Authors are advised to submit their bibtex database files. They are
%% requested to list a bibtex style file in the manuscript if they do
%% not want to use model1-num-names.bst.

%% References without bibTeX database:

% \begin{thebibliography}{00}

%% \bibitem must have the following form:
%%   \bibitem{key}...
%%

% \bibitem{}

% \end{thebibliography}

\end{document}